\documentstyle[12pt,aaspp4,flushrt]{article}

\begin{document}

\title{On the Evolution of the Globular Cluster
       Luminosity Function: The Differences}

\author{Oleg Y. Gnedin}

\affil{Princeton University Observatory, 
       Peyton Hall, Princeton, NJ~08544;
       ognedin@astro.princeton.edu}

\begin{abstract}
We consider the observational signature of the dynamical effects on the
luminosity function of globular clusters. For the three best studied
systems, in Milky Way, M31, and M87, there is a statistically
significant difference between the inner and outer population
of globular clusters. In all cases the inner clusters are on average
brighter than the outer clusters ($0.26 < \Delta m_0 < 0.84$) and
have a smaller dispersion in magnitudes ($0.04 < \Delta \sigma < 0.53$),
with the larger differences for the local, better observed samples.
The differences are of the type that would be expected if the inner population
had been depleted by tidal shocks.
The results suggest that the inner population suffers substantial
evolution from its initial distribution and cannot not be used as
a standard candle without correction for dynamical evolution.
\end{abstract}


\section{Introduction}
The turnover magnitude of the luminosity function of globular clusters
in external galaxies has been used as a standard candle for distance
measurements (cf \cite{Jetal:92}).
The method is usually applied to elliptical galaxies with the calibration
to the globular cluster system (GCS) in the Local Group.
Although it seems to be quite universal, the center of
the luminosity function can be expected to differ from galaxy to
galaxy. In fact, there is no consistent theoretical model that
predicts the same mean mass or luminosity for the {\it initial}
distribution of globular clusters (but see \cite{FR:85}, and \cite{VP:95}).
Recent work by \cite{GO:97} and \cite{MW:96} (1996a,b), which built
on less detailed analysis of \cite{AHO:88} and \cite{CW:90},
demonstrated that the GCS in our Galaxy suffers a substantial depletion
over the Hubble time due to a variety of dynamical processes, but primarily
the tidal bulge and
disk shocks and the enhanced evaporation of stars through the tidal
density cutoff. These effects vary with distance to the Galactic center
and are most pronounced in the inner part.
In this {\it Letter} we look for an observational evidence for the
imprints of the dynamical effects and address the question of universality
of the globular cluster luminosity function (GCLF).

If the dynamical effects are important, the inner population should
show a distribution different from that of the outer population
that presumably has the initial form (modulo internal two-body relaxation
process). For our Galaxy three-dimensional information is available, but
for others only projected radii are known. Therefore, the simplest approach
is to divide the observed sample into two halves, inner and outer, according
to the distance from the center of the galaxy.

The GCLF is conventionally fitted to a Gaussian function in magnitude,
$\phi(m) \propto \exp{(-(m-m_0)^2/2\sigma^2)}$. It provides a useful
measure of the overall shape and the parameters ($m_0, \sigma$)
of the GCS. We assume also that the two halves of the sample can be fit
by the same function, but with different parameters. This allows us
an easy way to compare the two populations. As an independent check that
does not use the assumption of normal distribution, we compare
the median points ($\mu$) of the two populations.

We use the maximum likelihood estimator (MLE) as a prime tool
to evaluate the parameters of the distribution and their uncertainties
(\cite{L:93}). \cite{HW:87} reviewed this method in detail and formulated
the necessary corrections for the magnitude-limited data.
We use their equations (10a--10c) to estimate the mean and dispersion
of the sample as well as the normalization coefficient, all of which are
allowed to vary. \cite{S:92} investigated a number of functional forms
to be applied in the ML analysis of the GCLF and found the Student's
$t$-distribution to be optimal. Since the deviation in the inferred parameters
as compared to the Gaussian is small, we chose the latter as it is simpler
and more commonly used.
In most studies up to now the turnover point and the dispersion of the GCLF
were determined by fitting a Gaussian to the histogram
of the magnitude distribution. This method implies binning of the data
and leads to a significant loss of information that becomes
especially important in case of a small sample (as in the Milky Way
and M31). We use it as a secondary estimator, mainly for consistency
with the previous results.

Monte-Carlo simulations were performed to establish the statistical
significance of the differences of the inner and outer populations.
The original sample was divided into two parts randomly, and for each such
realization the differences of the means, $\Delta m_0$, and of the medians,
$\Delta \mu$, were recorded. 10,000 of the random realizations constitute
a very nearly gaussian sample from which the probability of drawing
a given $\Delta m_0$ or $\Delta \mu$ is calculated.

\section{Milky Way}
The sample of the Galactic globular clusters is best studied and
essentially complete. We take the recent compilation by \cite{D:93} of
140 clusters with measured magnitudes and distances from the Galactic
center. Sorted by the distance sample is divided into two equal parts
at the boundary radius of $R_{1/2}=6.6$ kpc. Note that \cite{Z:93}
used a similar division to separate the old and young halo clusters.

The estimated parameters of the Milky Way GCLF are given in Table
\ref{tab:mw}. The mean of the whole distribution is at $M_V^0 = -7.17\pm 0.12$,
in accord with Secker's (1992) estimate of  $M_V^0 = -7.14$
(fitting the histogram, however, gives $-7.36\pm 0.17$; \cite{Hetal:91}).
The means of the two populations differ by almost 0.7 mag, and the medians
by 0.33 mag, with the inner part being systematically brighter. The outer
clusters also have much larger dispersion.
The result is statistically significant as given by the Monte-Carlo
probability of less than 0.5\% (Table \ref{tab:mw}) that the observed
$\Delta m_0$ occurs by chance. As expected, the dispersion of $\Delta m_0$
from the Monte-Carlo realizations is nearly the same as the MLE of the
standard error of $\Delta m_0$. The difference in medians is
smaller but nonetheless noticeable. We have checked that restriction
of the sample to the halo clusters (with [Fe/H] $<-0.8$; \cite{Z:85})
changes all parameters by much less than their standard errors.

\begin{figure}
\plotone{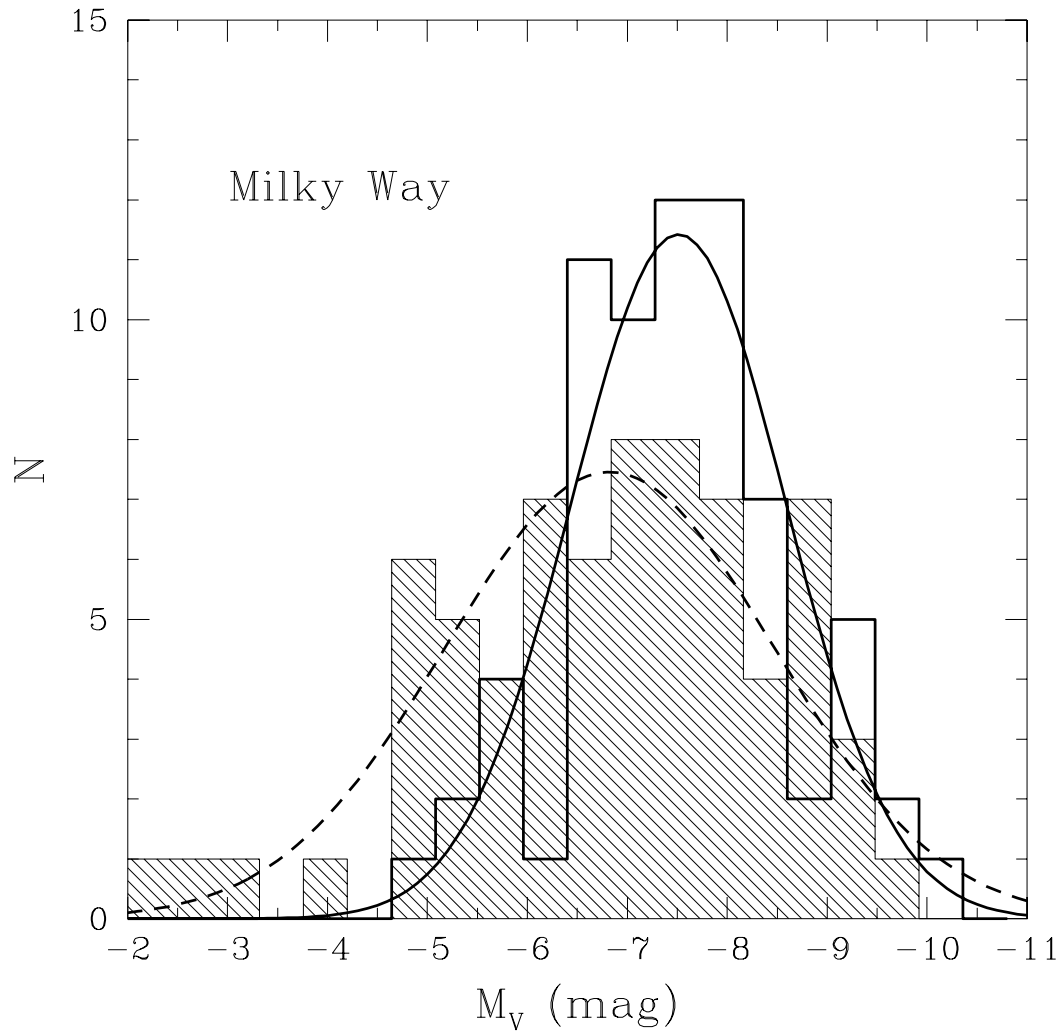}
\caption{Histograms for the inner (solid) and outer (shaded) populations
of the globular clusters in Milky Way. The two Gaussians with the
ML estimated parameters are superimposed on the histograms
for comparison. The solid line is for the inner sample, the dashed is for the
outer. The peak of the inner population is brighter by 0.7 mag, while the outer
population has considerably larger dispersion.\label{fig:mw}}
\end{figure}

The difference of the inner and outer populations is illustrated on
Figure \ref{fig:mw} that shows histograms for the two subsamples and our
fit obtained with the ML estimated parameters.
Fitting a Gaussian directly to the histograms
depends on the size of the bins. We found a slight variation within the errors
of the two means obtained this way for a range of bin sizes, $dm$,
from 0.3 to 0.5 mag.
Table \ref{tab:mw} shows the results for $dm = 0.45$ mag which are consistent
with the ML estimates. The dispersion of the outer sample is less well
constrained; note also the large standard error of its ML estimate.
The difference of the means is reduced to 0.5 mag, but still is
a two--sigma result.

\cite{KH:96} have recently undertaken a similar study and found
no difference in the centers of the inner and outer populations of the halo
clusters, though the dispersions were significantly different (and close to
our estimates). To facilitate a comparison with their results, we tried
to match the sample they used as close as possible; it was drawn from
the database maintained by W. Harris\footnotemark.
\footnotetext{\tt http://www.physics.mcmaster.ca/Globular.html}
The second half of Table \ref{tab:mw} shows that $\Delta m_0$ is still
significant at 1.5--sigma level when calculated using ML, but disappears when
inferred from the histogram fitting. For the latter we used the same bin size
(0.4 mag) as \cite{KH:96}. We doubt the trustiness of the latter estimate
based on binning of already very small sample ($N<50$). The lesser
statistical significance of the ML estimator of $\Delta m_0$ in this case
arises probably from the smaller size of the sample.

\section{M31}
The Andromeda has more globular clusters than Milky Way,
and the dust obscuration is less of a problem.
We use the most recent compilation of the GCS in M31
by J.-M. Perelmuter\footnotemark\ with 1 cluster rejected on basis
of its radial velocity (the kinematic data were kindly provided by J. Huchra).
\footnotetext{\tt http://www.astro.umontreal.ca/people/jperel/globs.html}
The sample becomes incomplete at $V \sim 17.5$, well beyond the peak
of the distribution. We considered, therefore, two cuts of the
original data at $V=18$ and $V=17.5$, respectively.

\begin{figure}
\plotone{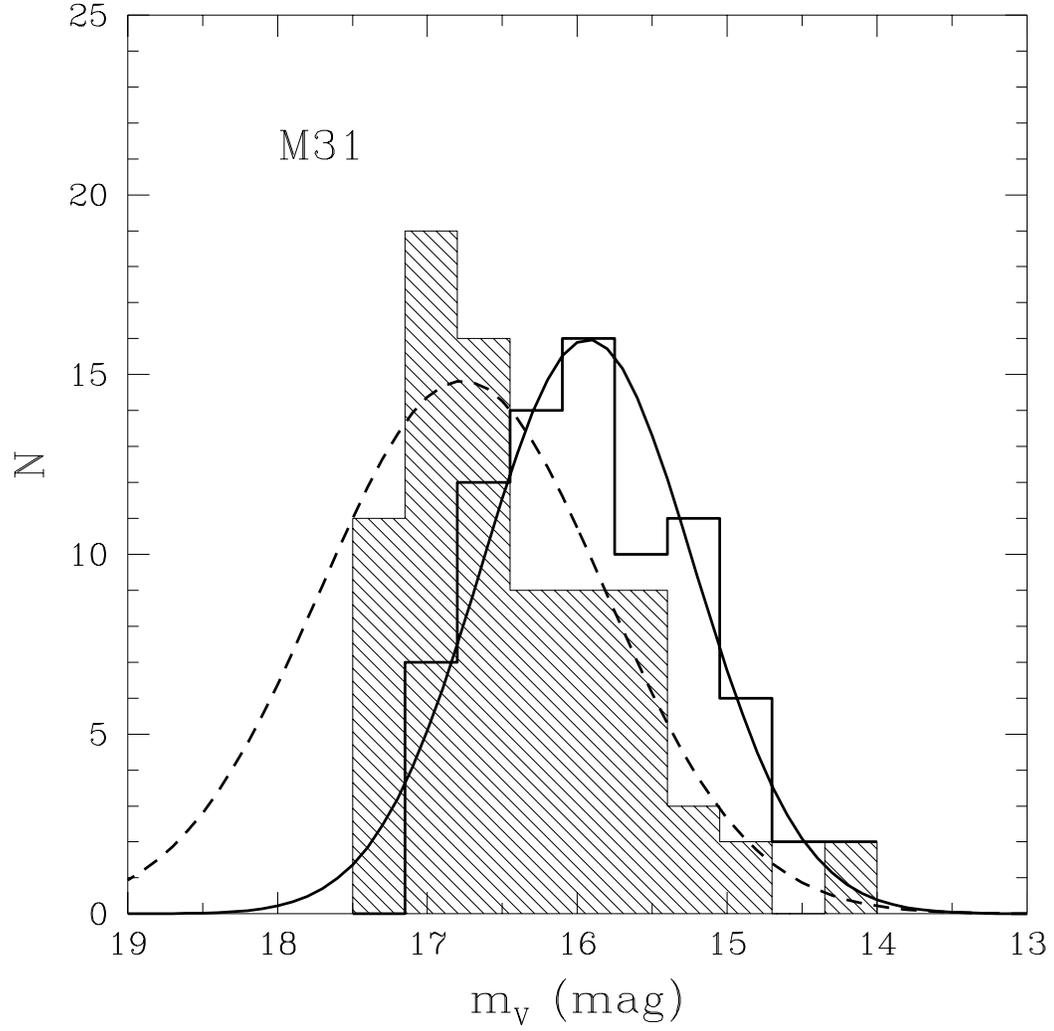}
\caption{Same as Figure \protect\ref{fig:mw}, but for the globular clusters
in M31. The sample is restricted to $V < 17.5$ due to incompleteness of
the data.\label{fig:m31}}
\end{figure}

Figure \ref{fig:m31} shows the histograms for the more conservative
second cut. The centers of the two populations are clearly different
with $\Delta m_0 \approx 0.8$. The difference in dispersions is
smaller than in Milky Way and is within the errors.

The ML estimates agree very well with the histogram fitting results
(see Table \ref{tab:m31}) for both cuts of the sample. The inner
population is consistently brighter than the outer with huge
statistical significance (7-sigma for $\Delta m_0$ and 6-sigma
for $\Delta \mu$). Thus there is little doubt that dynamical evolution
left a strong imprint in the inner parts of the galaxy. Relative
closeness of M31 makes the completeness limit much better off than
in M87, so the result is essentially unaffected by the observational
selection.

\section{M87}
The giant cD galaxy at the center of the Virgo cluster possesses the largest
sample of globular clusters observed around galaxies, although the completeness
is a severe problem. The most recent data available to us is from
\cite{MHH:94}. All sources have $V < 24$, and the completeness limit
is about $V \approx 23.5$.
Along with the real clusters there could be a contamination from
the foreground stars or background sources. We restricted the sample to a
range of radii ($1.'21 < R < 7'$) where the signal-to-noise ratio is greater
than unity.

\begin{figure}
\plotone{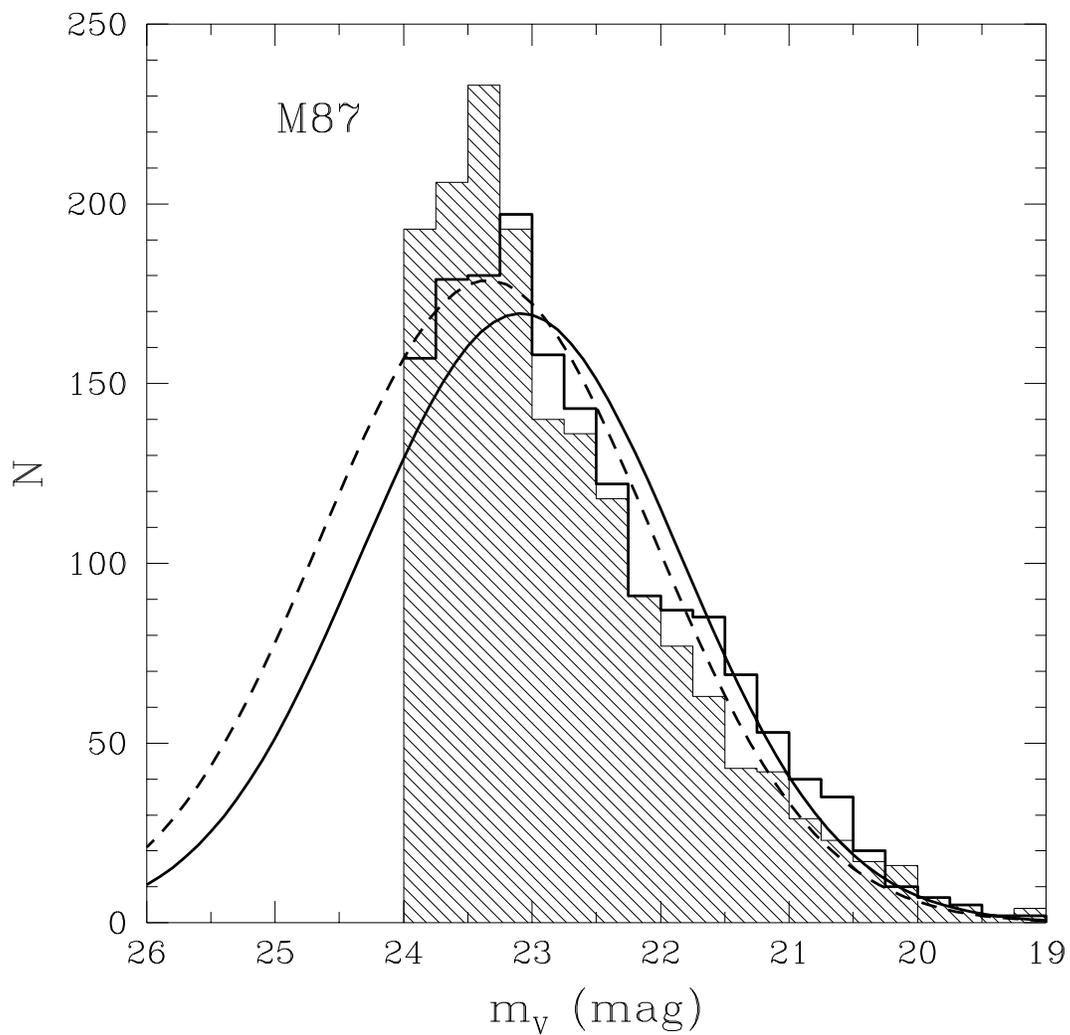}
\caption{Same as Figure \protect\ref{fig:mw}, but for the globular clusters
in M87. Only clusters with $V<24$ are shown, and the completeness limit
is about $V = 23.5$.\label{fig:m87}}
\end{figure}

Figure \ref{fig:m87} shows the histograms for the sample cut at $V=24$.
The inner and outer populations have similar distributions but there is
an apparent excess of the outer clusters at the faint end. Also, the inner
clusters prevail for $21 < V < 23$ (but not in the very bright end).

The MLE of the difference of the means is $\Delta m_0 = 0.26\pm 0.03$
(Table \ref{tab:m87}) and is formally very significant. These standard
errors should be taken with caution, however, as the systematic errors
due to incompleteness of the data are likely to be larger. For example,
the difference of the medians is only 0.16 mag. Similar results are
obtained when the sample is cut at $V=23.6$, i.e. very close to the peak
of the GCLF. In both cases statistical corrections applied to ML estimates
of $m_0$ and $\sigma$ are important. Therefore the results are somewhat
sensitive to the completeness limits. A deeper sample is required
for robust comparison of the inner and outer parts of the GCS.

Fitting the histograms becomes unstable because of the incompleteness.
Thus, even for the whole sample \cite{MHH:94} found that a broad range
of $m_0$ and $\sigma$ gives similar ``goodness'' ($\chi^2$) of the fit.
Our fits give fainter central magnitudes and broader dispersions, but
the scatter is very large. Also, different parameters were obtained for
different bin sizes. For this reason we consider the fits as unreliable.

\section{Conclusions}
We investigated the differences of the inner and outer populations
of globular clusters in Milky Way, M31, and M87.
There is a considerable difference in the turnover magnitudes,
$\Delta m_0 \approx 0.7-0.8$, for the two close and more complete
samples. Monte-Carlo simulations confirmed statistical significance
of the result. An independent check is provided by using the medians
instead of the means which is unaffected by the wings of the distribution.

In all three galaxies, the inner cluster population has brighter mean
magnitude and smaller dispersion. Presumably, the differences arise
due to dynamical processes operating most efficiently in the inner part
of the galaxies. The sign of the observed effects is as expected for a
depletion in the inner parts of the galaxies of low luminosity, low density
clusters. We will investigate the dynamical interpretation of the results
in a complementary study (\cite{OG:96}).

If the dynamical changes in the GCLF are significant, it cannot be used
directly as a standard candle. Special corrections
should be introduced that take into account the evolution.

\acknowledgements
I am greatly indebted to Prof. J. P. Ostriker who proposed this investigation
and provided constant support and encouragement. I would like to thank
R. Lupton for the numerous help with the statistics.
I am grateful to W. Harris, J. Huchra, D. McLaughlin, and J.-M. Perelmuter for
providing the data on the clusters in M31 and M87.
This project was supported in part by NSF grant AST 94-24416.


\begin{deluxetable}{lcccc}
\tablecaption{Globular Clusters in Milky Way\label{tab:mw}}
\tablecolumns{5}
\tablehead{\colhead{Sample} & \colhead{N} & \colhead{$m_0$ (mag)} &
           \colhead{$\sigma$ (mag)} & \colhead{$\mu$ (mag)}
           \\ \cline{1-5} \multicolumn{5}{c}{All clusters}}
\startdata
Total & 140 & $-7.17\pm 0.12$ & $1.43\pm 0.49$ & $-7.29\pm 0.09$ \nl
Inner &  70 & $-7.51\pm 0.13$ & $1.08\pm 0.44$ & $-7.40\pm 0.12$ \nl
Outer &  70 & $-6.82\pm 0.20$ & $1.65\pm 0.68$ & $-7.07\pm 0.22$ \nl
difference  &     & \phs $0.69\pm 0.24$ &            & \phs $0.33\pm 0.26$ \nl
probability\tablenotemark{a} & & $4.7\times 10^{-3}$ && $9.7\times 10^{-2}$ \nl
\cutinhead{Fit (bins 0.45 mag)}
Inner &     & $-7.48\pm 0.16$ & $1.04\pm 0.07$ \nl
Outer &     & $-6.99\pm 0.24$ & $1.44\pm 0.12$ \nl
\cutinhead{Comparison with \cite{KH:96}\tablenotemark{b}}
Total &  93 & $-7.16\pm 0.15$ & $1.40\pm 0.53$ & $-7.35\pm 0.11$ \nl
Inner &  49 & $-7.36\pm 0.15$ & $1.02\pm 0.46$ & $-7.43\pm 0.07$ \nl
Outer &  44 & $-6.94\pm 0.26$ & $1.71\pm 0.79$ & $-7.25\pm 0.32$ \nl
difference  &     & \phs $0.42\pm 0.30$ &            & \phs $0.18\pm 0.33$ \nl
probability\tablenotemark{a} & & 0.15   &            & 0.33 \nl
\cutinhead{Fit (bins 0.4 mag)}
Inner &     & $-7.38\pm 0.12$ & $0.64\pm 0.10$ \nl
Outer &     & $-7.39\pm 0.32$ & $1.39\pm 0.20$
\enddata
\tablenotetext{a}{Probability of the Monte-Carlo realizations that a given
result occurs by chance.}
\tablenotetext{b}{Sample constructed to match that used by \cite{KH:96}
from the database of W. Harris. The clusters have [Fe/H] $<-0.8$ and
lie within $1<R<85$ kpc from the Galactic center. The dividing boundary
is at $R_{1/2}=8$ kpc.}
\end{deluxetable} 

\begin{deluxetable}{lcccc}
\tablecaption{Globular clusters in M31\label{tab:m31}}
\tablecolumns{5}
\tablehead{\colhead{Sample} & \colhead{N} & \colhead{$m_0$ (mag)} &
           \colhead{$\sigma$ (mag)} & \colhead{$\mu$ (mag)}
           \\ \cline{1-5} \multicolumn{5}{c}{Clusters with $V < 18$}}
\startdata
Total & 168 & $16.29\pm 0.06$ & $0.88\pm 0.27$ & $16.32\pm 0.10$ \nl
Inner &  84 & $15.93\pm 0.08$ & $0.70\pm 0.27$ & $16.01\pm 0.07$ \nl
Outer &  84 & $16.71\pm 0.09$ & $0.94\pm 0.32$ & $16.68\pm 0.09$ \nl
difference & & \phn $0.78\pm 0.12$ &           & \phn $0.67\pm 0.11$ \nl
probability\tablenotemark{a} && $4.1\times 10^{-10}$ && $7.3\times 10^{-4}$ \nl
\cutinhead{Fit (bins 0.3 mag)}
Inner &     & $15.95\pm 0.10$ & $0.73\pm 0.14$ \nl
Outer &     & $16.70\pm 0.12$ & $0.78\pm 0.17$ \nl
\cutinhead{Clusters with $V < 17.5$}
Total & 160 & $16.32\pm 0.06$ & $0.88\pm 0.25$ & $16.25\pm 0.12$ \nl
Inner &  80 & $15.93\pm 0.08$ & $0.71\pm 0.27$ & $16.01\pm 0.07$ \nl
Outer &  80 & $16.76\pm 0.08$ & $0.95\pm 0.30$ & $16.63\pm 0.08$ \nl
difference & & \phn $0.84\pm 0.11$ &           & \phn $0.62\pm 0.11$ \nl
probability\tablenotemark{a} && $4.7\times 10^{-12}$ && $1.9\times 10^{-3}$ \nl
\cutinhead{Fit (bins 0.35 mag)}
Inner &     & $15.88\pm 0.09$ & $0.67\pm 0.16$ \nl
Outer &     & $16.83\pm 0.26$ & $0.93\pm 0.13$
\enddata
\tablenotetext{a}{Probability of the Monte-Carlo realizations that a given
result occurs by chance.}
\end{deluxetable} 

\begin{deluxetable}{lcccc}
\tablecaption{Globular Clusters in M87\label{tab:m87}}
\tablecolumns{5}
\tablehead{\colhead{Sample} & \colhead{N} & \colhead{$m_0$ (mag)} &
           \colhead{$\sigma$ (mag)} & \colhead{$\mu$ (mag)}
           \\ \cline{1-5} \multicolumn{5}{c}{Clusters with $V < 24$}}
\startdata
Total & 3287 & $23.22\pm 0.02$ & $1.26\pm 0.15$ & $22.92\pm 0.02$ \nl
Inner & 1644 & $23.09\pm 0.02$ & $1.24\pm 0.18$ & $22.84\pm 0.03$ \nl
Outer & 1643 & $23.35\pm 0.02$ & $1.28\pm 0.18$ & $23.01\pm 0.03$ \nl
difference & & \phn $0.26\pm 0.03$ &            & \phn $0.16\pm 0.05$ \nl
probability\tablenotemark{a} && $4.1\times 10^{-14}$ && $2.9\times 10^{-4}$ \nl
\cutinhead{Clusters with $V < 23.6$}
Total & 2718 & $23.06\pm 0.02$ & $1.21\pm 0.15$ & $22.69\pm 0.02$ \nl
Inner & 1359 & $22.93\pm 0.02$ & $1.18\pm 0.18$ & $22.63\pm 0.03$ \nl
Outer & 1359 & $23.19\pm 0.03$ & $1.24\pm 0.18$ & $22.73\pm 0.03$ \nl
difference & & \phn $0.26\pm 0.04$ &            & \phn $0.10\pm 0.04$ \nl
probability\tablenotemark{a} & & $2.4\times 10^{-13}$ & & $1.1\times 10^{-2}$
\enddata
\tablenotetext{a}{Probability of the Monte-Carlo realizations that a given
result occurs by chance.}
\end{deluxetable}

\end{document}